# Regulating Online Algorithmic Pricing: A Comparative Study of Privacy and Data Protection Laws in the EU and US

Zihao Li

**Abstract**

The emergence of big data, AI and machine learning has allowed sellers and online platforms to tailor pricing for customers in real-time. While online algorithmic pricing can increase efficiency, market welfare, and optimize pricing strategies for sellers and companies, it poses a threat to the fundamental values of privacy, digital autonomy, and non-discrimination, raising legal and ethical concerns. On both sides of the Atlantic, legislators have endeavoured to regulate online algorithmic pricing in different ways in the context of privacy and personal data protection. Represented by the GDPR, the EU adopts an omnibus approach to regulate algorithmic pricing and is supplemented by the Digital Service Act and the Digital Market Act. The US combines federal and state laws to regulate online algorithmic pricing and focuses on industrial regulations. Therefore, a comparative analysis of these legal frameworks is necessary to ascertain the effectiveness of these approaches. Taking a comparative approach, this working paper aims to explore how EU and US respective data protection and privacy laws address the issues posed by online algorithmic pricing. The paper evaluates whether the current legal regime is effective in protecting individuals against the perils of online algorithmic pricing in the EU and the US. It particularly analyses the new EU regulatory paradigm, the Digital Service Act (DSA) and the Digital Market Act (DMA), as supplementary mechanisms to the EU data protection law, in order to draw lessons for US privacy law and vice versa.

**Table of Contents**

## 1. Introduction

In the age of big data AI and machine learning, online platforms are able to generate users' digital profiles or infer their status by collecting and processing unprecedented volumes of data, which can strengthen platforms and sellers' ability to provide tailored and personalised prices to customers. For example, as early as 2001 Amazon was reportedly using cookies data to analyse customer behaviours, then selling products to different users for different prices.[1] Meanwhile, Staples.com is reported to frequently display lower pricing if competing shops were physically situated within around 20 miles of the customer's estimated location.[2] More recently, Uber is known to charge users with low-battery phones more, as they may be more desperate.[3] All these examples demonstrates that platforms and sellers have the capability to build users' digital footprints and employ algorithms to anticipate the price that end users will be willing to pay for products or services. Owing to the increasing problems of privacy invasion, loss of control of informational self-determination, and the inequality and unfairness caused by the uncertainty of legal regulation in this grey area, many professionals have called for legal interventions to counter the commonly applied practice of algorithmic pricing. For example, data privacy law is highly expected to regulate this area.

Several research already pointed out that the EU General Data Protection Regulation (GDPR) is capable to regulate online algorithmic pricing. For example, Zhao argues that the right to explanation empowered by data protection law is powerful enough to address the harms posed by online algorithmic pricing.[4] Furthermore, Esposito[5] and Steppe[6] respectively claims that the existing EU GDPR is already sufficient to regulate online algorithmic pricing. Many different toolkits, such as, right not to be subject to Automated Decision-Making (ADM) and right to object, can effectively govern such pricing strategies. While there are many innovative regulatory mechanisms, many overlook the fact that, despite appearing comprehensive and effective, the GDPR struggles to effectively regulate online algorithmic pricing primarily for two major reasons. Firstly, due to the ambiguity of the term "price discrimination" and "personalised pricing", many legal studies neglect affinity-based algorithmic pricing, which may bypass the EU GDPR.[7] Secondly, due to the ambiguity and legal uncertainty surrounding the definition of personal data, the practical effectiveness of the EU GDPR is greatly reduced.[8] For example, the GDPR is more effective to regulate the personal data-based algorithmic pricing than affinity-based algorithmic pricing.[9]

By comparing the US and EU's data protection and privacy regimes, this research endeavour to explore what lessons can EU and US data privacy regimes learn from each other in regulating online algorithmic pricing. It mainly argues that while the EU's comprehensive approach under the GDPR provides a robust model of consumer protection, it also exhibits significant gaps that could be mitigated by incorporating certain aspects of the US approach, particularly its flexible, sector-specific regulations. Meanwhile, US should also learn from the EU GDPR. The lack of a unified federal data protection law results in a patchwork of state and sector-specific regulations. This fragmentation leads to inconsistencies in enforcement and compliance requirements, making it challenging for companies to implement uniform privacy practices. Furthermore, this scattered approach fails to provide individuals with adequate data rights against online algorithmic pricing. More importantly, supplemented by the Digital Service Act (DSA) and Digital Market Act (DMA), the EU's digital regulation present a proactive and holistic approach to digital governance. These regulations provide a template for the US to potentially adapt, where similar comprehensive measures could be implemented to address the unique challenges posed by online algorithmic pricing. By adopting aspects of the EU's regulatory framework, the US could improve oversight of digital markets, increase transparency in algorithmic decision-making, and strengthen protections against the misuse of consumer data.

---

[1] Mark Ward, 'BBC News | BUSINESS | Amazon's Old Customers "Pay More"' (*BBC News*, 2000) <http://news.bbc.co.uk/1/hi/business/914691.stm> accessed 9 March 2021.
[2] Tim Worstall, 'Why Does Online Pricing Discriminate?' (*Forbes*, 2012) <https://www.forbes.com/sites/timworstall/2012/12/25/why-does-online-pricing-discriminate/> accessed 3 January 2022.
[3] Nicole Martin, 'Uber Charges More If They Think You're Willing To Pay More' (*Forbes*, 2019) <https://www.forbes.com/sites/nicolemartin1/2019/03/30/uber-charges-more-if-they-think-youre-willing-to-pay-more/?sh=7f9256c57365> accessed 3 November 2020.
[4] Zeyu Zhao, 'Algorithmic Personalized Pricing with the Right to Explanation' [2023] Journal of Competition Law & Economics <https://academic.oup.com/jcle/advance-article/doi/10.1093/joclec/nhad008/7259590> accessed 17 October 2023.
[5] Fabrizio Esposito, 'The GDPR Enshrines the Right to the Impersonal Price' (2022) 45 Computer Law & Security Review 105660.
[6] Richard Steppe, 'Online Price Discrimination and Personal Data: A General Data Protection Regulation Perspective' (2017) 33 Computer Law and Security Review 768.
[7] Zihao Li, 'Affinity-Based Algorithmic Pricing: A Dilemma for EU Data Protection Law' (2022) 46 Computer Law & Security Review 1, 4–6.
[8] ibid 7–9.
[9] ibid 4–6.

Against this backdrop, this working paper compares the regulatory approaches of the EU and the US in the regulation of online algorithmic pricing, exploring potential strategies for mutual improvement. After providing background and defining online algorithmic pricing through a review of legal, economic and technical literature, this paper points out the limitations and gaps in existing research concerning online algorithmic pricing. Building on this review, this paper further scrutinises the benefits and harms of online algorithmic pricing, critiquing both the laissez-faire and prohibition approach. The benefit of online algorithmic pricing includes increased transaction efficiency, enhanced market total welfare, and respect for business freedom, while the harms are more far-ranging. According to different form of online algorithmic pricing, the harm includes privacy invasion, derogation of informational autonomy and equality, and group privacy interest infringement. By examining both the EU and US data privacy laws' scope, *ex-ante* and *ex-post*, this paper identifies the drawbacks and advantages of each, endeavouring to propose the potential strategies for mutual improvement. Additionally, in the context of the new EU regulatory paradigm in online platform, this paper will also explore what lessons can US learn from the EU's experience.

## 2. Online Algorithmic Pricing: Benefits and Harms

As has been discussed by several economists, retailers are strongly incentivised to apply online algorithmic pricing due to its potentially lucrative nature.[10] The advent of big data and predictive algorithms also carries strong economic attributes. When companies have more information on consumer preferences, they also have more opportunities to extract consumer surplus through online algorithmic pricing.[11] Ideally, online algorithmic pricing can increase the consumer surplus and benefit both consumers and companies.[12] However, online algorithmic pricing also risks harming both consumers, the entire market and even society in certain circumstances. Therefore, it is necessary to explore its benefits and harms in line with different forms of online algorithmic pricing.

---

[10] Antonio Capobianco and Pedro Gonzaga, 'Competition Challenges of Big Data: Algorithmic Collusion, Personalised Pricing and Privacy', *Legal Challenges of Big Data* (Edward Elgar Publishing 2020) <https://www.elgaronline.com/view/edcoll/9781788976213/9781788976213.00008.xml>.
[11] A consumer surplus happens when the price consumers pay for a product or service is lower than the price they are actually willing to pay. It is therefore a measure of the additional benefit that consumers receive because they are paying less for something than they were willing to pay.
[12] Capobianco and Gonzaga (n 10) 57; Frederik Zuiderveen Borgesius and Joost Poort, 'Online Price Discrimination and EU Data Privacy Law' (2017) 40 Journal of Consumer Policy 347, 353–355.

## 2.1. Benefits of Online Algorithmic Pricing

As has already been demonstrated by several researchers, a good argument can be made in favour of price discrimination from a welfare economic perspective, since it can benefit both buyers and sellers. Although people will always complain that they suffer from price discrimination, some also benefit from discount offers too. Advocates of algorithmic pricing claim that it can benefit buyers on low incomes by offering them discounts, which helps to increase social efficiency.[13] By doing so, it could also lower the cost threshold for certain products that are beyond the financial reach of specific groups, thereby benefiting more individuals with access to higher-quality products. Similarly, the UK Office of Fair Trading (OFT) has failed to "conclude whether, in general, online personalised pricing is harmful or beneficial to consumers,"[14] because it is impossible to generalise the economic effects of online personalised pricing as the welfare outcomes are too highly dependent on different marketing variables.

Online algorithmic pricing can also benefit sellers due to its profitable nature. This pricing strategy can enhance overall market welfare by charging more from those with a higher willingness to pay, while offering lower prices to those who normally could not afford the product, provided the price still covers the sellers' costs.[15]

The above statement sounds hypocritical from a legal and individual standpoint, but the benefits of algorithmic pricing should not simply be neglected. Therefore, neither laissez-faire nor prohibition approach should be adapted. This does not mean that an online platform has the right to conduct e-commerce in any way it wishes. Instead, it implies that if algorithmic pricing can be done in a fairer way by giving individuals more control over their data, then the personalisation of online targeting could be a win-win situation. Therefore, simply blocking algorithmic pricing is not the right choice. Again, the debate returns to the question of the user's autonomy.

## 2.2. Harms of Online Algorithmic Pricing

Indeed, aside from certain advantages discussed in above, online algorithmic pricing has already raised risks. Some automated uses of personal data are potentially harmful to individuals and society as they breach several fundamental rights, and therefore require legal intervention. Such threats are closely connected with the taxonomy of online algorithmic pricing because different types of online algorithmic pricing may involve different values and risks.

As online algorithmic pricing often involves the increasingly pervasive presence of online tracking, profiling, and targeting, the online algorithmic pricing will first bring its potential risks to privacy at the individual level. Online tracking, a cornerstone of this pricing strategy, entails the continuous monitoring of users' digital footprints across websites and applications.[16] This extensive data collection is fundamental to the operation of algorithmic pricing models, as it feeds the algorithms with the necessary inputs to adjust prices dynamically based on user behaviour and characteristics.[17] However, such data collection often occurs without explicit consent from the individuals being tracked, thereby encroaching upon their privacy. Moreover, the sheer volume and detail of data collected, including locations, device information, and browsing history, can lead to significant privacy breaches if not adequately protected.

---

[13] Capobianco and Gonzaga (n 10) 49–55; Akiva A Miller, 'What Do We Worry About When We Worry About Price Discrimination? The Law and Ethics of Using Personal Information for Pricing' (2014) 19 Journal of Technology Law and Policy 41, 41.
[14] UK Office of Fair Trading, Patrick Coen and Natalie Timan, 'The Economics of Online Personalised Pricing' [2013] Office of Fair Trading Working Paper 97, 11; UK Competition and Markets Authority, 'Pricing Algorithms: Economic Working Paper on the Use of Algorithms to Facilitate Collusion and Personalised Pricing' [2018] UK Competition and Markets Authority Working Paper.
[15] Marco Botta and Klaus Wiedemann, 'To Discriminate or Not to Discriminate? Personalised Pricing in Online Markets as Exploitative Abuse of Dominance' (2020) 50 European Journal of Law and Economics 381, 388.

Further exacerbating privacy concerns are the profiling and targeting practices used in conjunction with algorithmic pricing. Profiling involves the aggregation and analysis of collected data to create detailed consumer profiles, which are then used to predict purchasing behaviours and set individualized prices. This not only raises issues of consent and transparency but also highlights the risk of discriminatory outcomes, where prices are manipulated based on personal characteristics inferred from the data. Additionally, targeting strategies that use these profiles to tailor marketing and pricing in real-time can perpetuate inequalities and potentially exploit vulnerable features of individuals.[18] Together, these practices highlight the urgent need for robust regulatory frameworks that enforce transparency, consent, and fairness in the application of online algorithmic pricing. As the underpinning principles of data protection, the values of fairness and human dignity can be undermined by online algorithmic pricing.

Further to these individual concerns, the new collective privacy interest, namely group privacy is also at risk.[19] Such concerns arise when data from individuals are aggregated to make assumptions and decisions about groups of people, such as those sharing certain demographic traits or behaviour patterns.[20] This form of privacy is particularly threatened in the context of algorithmic pricing, where companies may use aggregated data to set different price points for specific communities or demographic groups. Such practices can inadvertently lead to discriminatory effect and social stratification, where entire groups may be disadvantaged or unfairly targeted based on the collective data profile. This not only undermines individual privacy but also poses broader social risks by reinforcing existing inequalities and potentially introducing new forms of bias. Thus, protecting group privacy requires a careful reconsideration of how data is used in aggregate, and how these uses comply with ethical standards and legal norms concerning privacy protection, fairness and discrimination.

## 3. Regulating OAP in the EU data protection law

### 3.1. Scope of the GDPR in the Context of Online Algorithmic Pricing

---

[16] Jiahong Chen, *Regulating Online Behavioural Advertising Through Data Protection Law* (Edward Elgar Publishing 2021) <https://www.elgaronline.com/view/9781839108297.xml>.

[17] Le Chen, Alan Mislove and Christo Wilson, 'An Empirical Analysis of Algorithmic Pricing on Amazon Marketplace', *Proceedings of the 25th International Conference on World Wide Web* (International World Wide Web Conferences Steering Committee 2016) 1346–1348 <https://dl.acm.org/doi/10.1145/2872427.2883089> accessed 8 January 2023.

[18] Sandra Wachter, 'Affinity Profiling and Discrimination by Association in Online Behavioural Advertising' (2020) 35 Berkeley Technology Law Journal 1, 41–42.

[19] Brent Mittelstadt, 'From Individual to Group Privacy in Big Data Analytics' (2017) 30 Philosophy and Technology 475.

[20] Linnet Taylor, Luciano Floridi and Bart van der Sloot, *Group Privacy: New Challenges of Data Technologies* (Springer International Publishing 2017).

As long as the data collection and processing involve personal data, then the EU GDPR can intervene to regulate, and individuals can exercise their various data right to fight against algorithmic pricing. Although many legal scholars have argued that the GDPR should undoubtedly apply to online algorithmic pricing,[21] affinity-based online algorithmic pricing is highly likely to bypass the GDPR in reality. In fact, three types of data are stipulated by the GDPR: personal data, pseudonymous data, and anonymous data.[22] However, the GDPR can only cover personal data and pseudonymous data, while anonymous data is largely ignored. Article 4(1) of the GDPR defines personal data as "(i) any information (ii) relating to (iii) an identified or identifiable natural person", which can be further divided into personal data and pseudonymous data.

If the algorithmic pricing heavily relies on the data that can directly or indirectly identify individual, such as IP addresses, cookies, and geo-location data, then it shall generally satisfy the definition of personal data. For example, Amazon utilizes cookies to analyse customer behaviours, which influences different price bands. Here, IP addresses and similar data serve as unique identifiers, allowing for user tracking, as recognized by the CJEU in the Breyer case.[23] Consequently, personal data-based algorithmic pricing employing such technologies clearly falls under GDPR scope. However, the case for affinity-based pricing, which does not rely on personal identifiers, warrants further legal exploration. Because such affinity data may fall into the scope of anonymous data, but the former can still reveal certain status information for making pricing. First, some models infer demographic attributes from anonymous data, enhancing with current big data and AI capabilities, sensitive information like location and age can be deduced from users' communication patterns without revealing their identities. Second, data profiling through context-aware algorithms allows platforms to set prices based on inferred social tags and preferences without direct personal data processing.[24] Both affinity data and transient data processing technologies may bypass GDPR requirements by design. Third, affinity data is normally de-identifiable, used to infer private information, categorizes users for differential pricing based on group characteristics rather than individual identities. Even if this data can be classified as personal data, the GDPR exempts this type of data from several fundamental data rights.

However, there is still inconsistency in classifying such affinity data and inference data between Art. 29 WP and the CJEU, particularly regarding data impacting personal rights, underscores significant legal ambiguity. For example, the CJEU in cases like *YS, M and S*, and *Nowak* has indicated that not all data influencing personal decisions or private life should be considered personal data, especially when such data is derived from legal analyses or assessments.[25] The CJEU also restricts the remit of data protection law to not assessing the accuracy of a decision or the process of decision-making, as highlighted in *European Commission v. Bavarian Lager*.[26] This limitation affects data subjects' rights to challenge or rectify incorrect inferences made from their data, which is particularly pertinent in algorithmic pricing. Yet, such a flawed or even incorrect inference data is sufficient for online algorithmic pricing, because of the high tolerance of mistakes in this practice.[27] Indeed, the gap between the inferred results and users' real situations does not affect price setting, and such criteria are more likely to conversely cause algorithmic bias and discrimination, widening the possibility (and range of victims) of discriminatory behaviours.

---

[21] Benjamin Wong, 'Online Personalised Pricing as Prohibited Automated Decision-Making under Article 22 GDPR: A Sceptical View' (2021) 30 Information & Communications Technology Law 193, 2–3; Steppe (n 6) 772–775; Zuiderveen Borgesius and Poort (n 12) 356–358.
[22] Paul Voigt and Axel von dem Bussche, *The EU General Data Protection Regulation (GDPR) A Practical Guide* (Springer International Publishing 2017).
[23] Case C-582/14, Patrick Breyer v. Bundesrepublik Deutschland [2016] ECLI:EU:C:2016:779.
[24] Lan Zhang and others, 'Context-Aware Recommendation System Using Graph-Based Behaviours Analysis' (2021) 30 Journal of Systems Science and Systems Engineering 482, 482–483.
[25] Cases C–141 & 372/12, YS, M and S v. Minister voor Immigratie, Integratie en Asiel, 2014 E.C.R. I-2081.
[26] Case C-28/08 P, European Comm'n v. Bavarian Lager, 2010 E.C.R. I-6055.

In summary, the legal framework surrounding personal data-based algorithmic pricing is well-established under GDPR. However, the nature of affinity-based algorithmic pricing remains unclear, as it often does not meet the criteria for identifying specific users. Consequently, while individuals lack control and protection against such pricing mechanisms, these practices can still intrude on personal privacy and lead to discrimination. The ongoing legal ambiguity between Art. 29 WP and CJEU jurisprudence further complicates the enforcement of data protection laws in this context.

### 3.2. *Ex-ante* Measures of the GDPR

The GDPR imposes several *ex-ante* obligations designed to regulate commercial practices like online algorithmic pricing, fundamentally aimed at ensuring transparency, accountability, and fairness in the use of personal data. The GDPR mandates the establishment of Data Protection Impact Assessments (DPIAs) as *ex-ante* measures for operations likely to result in high risks to the rights and freedoms of individuals, which include complex algorithmic pricing schemes. These assessments require companies to evaluate and mitigate risks associated with data processing activities before they commence. It requires data controllers to systematically and comprehensively evaluate how personal data is used in their pricing algorithms, the necessity and proportionality of such processing, and the risks to individuals' privacy. In the context of personal data-based algorithmic pricing,[28] DPIAs serve as an essential tool to ensure transparency and accountability. It forces organisations to map out the flow of personal data, assess the impact of their processing activities, and identify measures to mitigate any potential privacy risks. This includes examining how variables such as IP address, tracking data, or past purchasing behavioural data are used to influence pricing, and whether these practices could lead to discrimination or unfair treatment of individuals or groups.

Additionally, Privacy by Design (PbD) is another *ex-ante* foundational principle integrated into the GDPR that emphasizes the inclusion of data protection from the inception of technology systems and business practices.[29] This principle mandates that privacy and data protection are not merely added as an afterthought, but are embedded within the infrastructure of IT systems and business operations.[30] In terms of online algorithmic pricing, PbD requires that data protection measures are built into the algorithms and the surrounding processes used to determine pricing strategies. Indeed, the employment of techniques such as data minimization, where only the necessary amount of personal data is processed, and pseudonymization, which can help mask user identities even if data breaches occur, is helpful. These techniques are effective when sellers use personal data-based algorithmic pricing. By embedding these protective measures from the beginning, businesses can only collect and process users' personal data for their specific purposes.

---

[27] Peter Seele and others, 'Mapping the Ethicality of Algorithmic Pricing: A Review of Dynamic and Personalized Pricing' (2021) 170 Journal of Business Ethics 697.
[28] Li, 'Affinity-Based Algorithmic Pricing: A Dilemma for EU Data Protection Law' (n 7) 7.
[29] Fei Bu and others, '"Privacy by Design" Implementation: Information System Engineers' Perspective' (2020) 53 International Journal of Information Management 102124.
[30] Zihao Li, 'Standardisation of Blockchain and Distributed Ledger Technologies—A Legal Voice from the Data Protection Law Perspective' in Kai Jakobs (ed), *25th EURAS Annual Standardisation Conference – Standards for Digital Transformation: Blockchain and Innovation* (The European Academy for Standardisation 2020).

While the both DPIA and PbD offers a robust framework for embedding privacy considerations into business practices, its application has limitations, particularly when dealing with affinity data, which does not necessarily identify specific individuals but is used to influence group behaviours and decisions, such as affinity-based algorithmic pricing. One significant limitation of PbD in this context is the challenge of defining and enforcing privacy standards for data that is not directly personal but still capable of affecting individual privacy when aggregated. Affinity data, which might include information about device use, browsing patterns, or consumer group preferences, can be used to make inferences about user behaviours or preferences at a group level rather than individual level. This type of data often falls into a grey area under GDPR because it does not always meet the strict definition of "personal data."[31] Both DPIA and PbD assumes that all data can be clearly categorised and protected according to its sensitivity and identifiability. However, when dealing with affinity data, it is challenging to apply these principles because the indirect consequences of data aggregation are difficult to predict and manage. For instance, while individual data points may not identify a person, their compilation through sophisticated algorithms can lead to conclusions that closely reflect personal identities or characteristics, potentially leading to privacy invasions or discriminatory outcomes.

Moreover, PbD's effectiveness is contingent on the proactive measures taken during the design phase of systems and processes. If the potential for indirect identification or privacy impacts is underestimated, or if the technology evolves in ways not anticipated by initial designs, the privacy safeguards may prove inadequate. This highlights another limitation: the dynamic nature of data analytics and machine learning can outpace the static data protection measures envisioned at the time of a system's design. This discrepancy amplifies a dilemma where the design of data systems and algorithms is typically based on existing data sets and usage scenarios, which are expected to guide the system's future behaviour. However, as these systems learn and adapt, they can develop new functionalities or extend their data-processing capabilities beyond their original scope. This evolutionary capability can lead to the creation of new privacy risks that were neither anticipated nor addressed by the initial DPIA or PbD safeguards. Such challenges are more evident in the context of affinity-based algorithmic pricing, where data that does not directly identify individuals is utilized to influence market behaviours and individual purchasing decisions. In these scenarios, the dynamic nature of data analytics and machine learning can significantly outpace the static data protection measures originally envisioned. As algorithms evolve, they can begin to discern patterns and draw inferences from aggregated, non-personal data, transforming it into a tool that, while not identifying individuals explicitly, significantly impacts consumer privacy and market fairness.

In summary, while both DPIA and PbD is a critical component of the GDPR, its application to affinity data presents distinct challenges. These include difficulties in applying data protection principles to deidentification data, anticipating the long-term implications of aggregated data use, and maintaining effective oversight and enforcement as technology and data use practices evolve.

### 3.3. *Ex-post* Measures of the GDPR

---

[31] Li, 'Affinity-Based Algorithmic Pricing: A Dilemma for EU Data Protection Law' (n 7) 8–11.

Apart from *ex-ante* measures, the GDPR also establishes several *ex-post* remedies for individuals against online algorithmic pricing. These remedies are essential for offering recourse and addressing concerns that arise from the online pricing algorithms. These remedies, pivotal in the realm of data rights, include the right to access, right to rectification, right to erasure, right to object, and the rights against automated decision-making. Each of these rights plays a crucial role in countering potential misuses of algorithmic pricing and ensuring data subject autonomy.

Art. 15 GDPR stipulates the right to access, which grants individuals the ability to obtain a copy of their personal data that is being processed by a data controller. This right is foundational in the context of algorithmic pricing as it allows data subjects to inspect the data that informs their pricing category and to understand the decision-making logic behind the algorithms. As some scholars argue, this right constitutes the right to explanation.[32] It enables individuals to verify the accuracy of the data and assess whether the data processing is being done fairly and lawfully. By providing transparency, this right empowers consumers to challenge decisions that may be based on flawed or discriminatory algorithms, thus promoting accountability among businesses using algorithmic pricing models. However, it is worth noting that sellers only need to provide the purposes of the data processing, the categories of personal data, and the source of the data when individuals exercise this right. It also remains uncertain that if such right can apply to inferred and derived data on profiling, as Recital 63 highlights that exercising the right to access should not adversely impact trade secrets and intellectual property, though the former is often considered as a trade secret protected by sellers.[33]

Additionally, Art. 16 and Art. 17 establish the right to rectification and erasure respectively. The right to rectification ensures that individuals can correct false or incomplete data that concerns them.[34] This right is critical in algorithmic pricing scenarios where pricing decisions might be based on inaccurate or outdated personal data. If left uncorrected, such data can lead to unfair or biased pricing outcomes. Enabling individuals to update their data ensures that decisions made by automated systems are based on the most accurate and current information, thereby fostering fairness in digital commerce practices. Similarly, the right to erasure, often referred to as the 'right to be forgotten', enables individuals to demand the deletion of their personal data when it is no longer necessary for the purposes it was collected for, or when the individual withdraws consent. This right is particularly significant for mitigating the impact of historical purchasing behaviours on future algorithmic pricing strategies, as it allows for the removal of data from profiles utilized by pricing algorithms. However, in the context of algorithmic pricing, the rights of rectification and erasure are only marginally useful. Firstly, data subjects must know they have been charged through algorithmic profiling. Secondly, the algorithmic price has to be inaccurate or based on inaccurate personal data. However, from the analysis above regarding the right to access personal data, it is clear that data subjects do not have sufficient information to know the type of personal data used in algorithmic pricing. Therefore, huge information asymmetry between data subjects and controllers exists, impeding the exercise of the right to rectify.

---

[32] Zhao (n 4).
[33] Gianclaudio Malgieri, 'Trade Secrets v Personal Data: A Possible Solution for Balancing Rights' (2016) 6 International Data Privacy Law 102.
[34] Voigt and von dem Bussche (n 22).

Moreover, right to object, stipulated by Art. 21 GDPR, permits individuals to object to the processing of their personal data, including profiling, especially when the processing is for direct marketing purposes. Unless the controllers can demonstrate "compelling legitimate grounds" for the process that override the rights and the freedom of data subjects, it is necessary for controllers to stop the processing. In the context of algorithmic pricing, this right is significant as it enables individuals to refuse the use of their personal data for creating personalized pricing strategies. As researcher argue, this right could empower individual receive impersonal pricing.[35] It provides a means for consumers to resist being targeted by dynamic pricing models that may lead to price discrimination or manipulation based on personal data profiles. However, Art. 21 stipulates that controllers' compelling legitimate grounds can provide an exemption from this right. However, a definition of those compelling legitimate grounds is absent. Similarly, the extent to which a legitimate ground can override the right and interests of data subjects remains unclear, which makes it questionable whether the sellers' fundamental right and freedom to conduct their business can constitute the compelling legitimate grounds required.[36] In addition, although it is undisputed that Article 21 can be applied to direct online marketing, the GDPR lacks a definition of direct marketing.[37] As a result, whether online algorithmic pricing can be seen as a kind of online direct marketing is doubtful. According to the proposed e-Privacy Regulation, Article 4(3)(f) defines direct marketing communications as "any form of advertising, whether written or oral, sent to one or more identified or identifiable end-users of electronic communications services",[38] which narrows down the scope of direct marketing to the advertising level for either identified or identifiable end-users.[39] Accordingly, online algorithmic pricing should not be included in direct marketing, which in turn should not be covered by this Article. Consequently, Article 21 can only provide limited assistance in the context of algorithmic pricing, which means that its regulation of algorithmic pricing still needs further observation.

Article 22 provides safeguards against decisions made solely on the basis of automated processing, including profiling, that have legal or similarly significant effects on individuals.[40] This provision is crucial in regulating algorithmic pricing that excludes human oversight. It mandates that individuals have the right to challenge and obtain human intervention for decisions derived from automated processes, ensuring that any significant decisions, such as those affecting pricing or access to services, are fair and can be contested. This right addresses the opacity of fully automated decisions and helps mitigate risks related to bias, ensuring that algorithmic decisions are subject to scrutiny and accountability. However, Article 22 of the GDPR faces notable limitations, particularly regarding the scope of its applicability to solely automated decisions.[41] This term is narrowly defined, allowing organizations to circumvent these rules by introducing nominal human involvement, which may not substantively affect the outcome but does exempt the process from this article's constraints. Additionally, Article 22 only applies to decisions that have "legal or similarly significant effects" on individuals, yet the GDPR does not clearly define these terms. This ambiguity leads to variability in interpretation, potentially excluding some automated decisions, such as those involving algorithmic pricing, from the protections intended by this provision.

---

[35] Esposito (n 5).
[36] Li, 'Affinity-Based Algorithmic Pricing: A Dilemma for EU Data Protection Law' (n 7) 11–14.
[37] Sandra Wachter and Brent Mittelstadt, 'A Right to Reasonable Inferences: Re-Thinking Data Protection Law in the Age of Big Data and AI' (2019) 2 Columbia Business Law Review 443, 52–53.
[38] Proposal for a Regulation of the European Parliament and of the Council concerning the respect for private life and the protection of personal data in electronic communications and repealing Directive 2002/58/EC 2017.
[39] Zihao Li, 'Digital Advertising and EU Digital Regulation', *Digital Advertising Evolution* (Routledge 2024).
[40] Steppe (n 6) 783–784.
[41] Zuiderveen Borgesius and Poort (n 12) 361–362.

As aforementioned, these ex-post remedies under the GDPR not only provide individuals with tools to manage how their personal data is used but also impose a regulatory check on companies using algorithmic pricing models. By requiring these firms to adhere to principles of transparency, accuracy,[42] and fairness, the GDPR helps mitigate the potential negative impacts of algorithmic pricing. However, the effectiveness of these remedies often depends on the interpretations of essential terms, proactive enforcement of rights by data subjects and the oversight of regulatory authorities. Challenges such as identifying the specific data used in algorithms, deidentification dilemma and the technical complexity of how such data is processed remain significant barriers. Therefore, while the GDPR lays a solid foundation for protecting consumer rights in the digital age, continuous monitoring, enforcement, and potentially further regulatory guidance are needed to address the evolving nature of algorithmic pricing and their implications for individual rights.

## 4. Regulating OAP in the US data privacy law

### 4.1. Scope of the US data privacy law in the Context of Online Algorithmic Pricing

In contrast, the US data privacy law is a completely different framework. Based on the conception of data marketability, the US data privacy law mainly focuses on the protection in a marketplace marked by deception and unfairness and safeguarding of individuals participating in the digital market. In this context, individuals are also referred to as "privacy consumers" in different federal or state privacy legislations. In this view, the individual is often regarded as a trader of their personal commodity, such as their behavioural data.[43] According to this line of reasoning, the individual partakes in the digital marketplace as a participant, sharing both the benefits and losses within the entire realm of the digital economy. Even though such mindset is completely different than the EU, protecting individual against online algorithmic pricing is extremely essential as this commercial practice often involves unfair and discriminatory trade, which could impede the interests of individuals and hinder the prosperity of the whole digital market. For instance,

The second feature of the US data privacy law is that the whole regulation is based on the scattered and fragmented sector-based framework. This framework eschews a comprehensive federal data privacy law in favour of disparate statutes tailored to specific industries or data types. For instance, the Federal Trade Commission Act (FTC Act) is one of the primary federal tools for regulating algorithmic pricing, especially under Section 5, which prohibits unfair or deceptive acts or practices in commerce. The FTC has authority to take action against companies that use algorithmic pricing in ways that are deceptive (e.g., not disclosing how prices are determined) or unfair (e.g., discriminatory pricing that could harm consumers). The Health Insurance Portability and Accountability Act (HIPAA) rigorously protects health information, while the Gramm-Leach-Bliley Act (GLBA) specifically addresses financial data. This results in a variegated regulatory landscape for algorithmic pricing, where the applicability and stringency of privacy regulations can differ markedly depending on the sector. Such sector-based framework also offers a different angle of oversight based on its foundational objectives, including consumer protection, anti-discrimination, data privacy, or competition law. Therefore, it is hard to distinguish a single data privacy perspective. Therefore, in the following analysis, this research will mainly focus on the main industrial regulations that can cover online algorithmic pricing.

---

[42] Zihao Li, Weiwei Yi and Jiahong Chen, 'Accuracy Paradox in Large Language Models: Regulating Hallucination Risks in Generative AI'.
[43] Ruben De Bruin, 'A Comparative Analysis of the EU and US Data Privacy Regimes and the Potential for Convergence' [2022] Hastings Sci. & Tech. L.J. 130 <https://www.ssrn.com/abstract=4251540> accessed 30 April 2024.

The third feature of the US data privacy law is its state-led data privacy legislations, which arise due to the absence of a unified federal privacy law. This decentralized approach allows individual states to fill the regulatory void. Notable examples include California's Consumer Privacy Act (CCPA), which offers some of the most stringent data privacy protections in the country, and Virginia's Consumer Data Protection Act (CDPA), which similarly regulates the processing and control of personal data.[44] Regardless of which state a company is located in, the rights the laws provide apply only to people who live in these states. Such fragmentation not only increases operational complexity and costs for businesses but also creates uneven protections for consumers depending on their state of residence. As this research examines EU data protection law at the EU level, to maintain a balanced and equitable comparison, the following comparison will examine US data protection laws at the federal level. By concentrating on federal laws in the US, this research avoids the complexity and limited application of the US's patchwork of state laws, allowing for a more straightforward analysis.

Based on above analysis, several US regulations can govern online algorithmic pricing. The FTC Act plays a central role in regulating online algorithmic pricing through its prohibition of unfair and deceptive practices. According to the Section 5 of the FTC Act, if the algorithmic pricing strategy is deemed unfair and causes substantial consumer harm that consumers cannot reasonably avoid themselves and that is not outweighed by countervailing benefits to consumers or competition, then the FTC authority can intervene.[45] For instance, if an online retailer uses an algorithmic pricing model that discriminates against certain consumer segments by charging them higher prices based on their zip code or browsing history without disclosing the fact, such practices could be scrutinized under the unfairness doctrine. Moreover, if companies use algorithmic pricing and mislead consumers about how prices are determined and being personalised, such practices could be challenged as deceptive under the FTC Act. While the application scope seems much broader than the EU GDPR, there are significant limitations to its effectiveness. One of the major hurdles under the unfairness standard is the requirement to prove that a practice causes substantial consumer harm. In the context of algorithmic pricing, quantifying this harm and directly attributing it to a specific pricing algorithm can be exceedingly difficult, especially when the criteria and data inputs used by the algorithm are opaque.

Apart from the FTC Act, Americans with Disabilities Act (ADA) could intervene if an algorithmic pricing model inadvertently leads to discriminatory pricing against people with disabilities. This can occur in scenarios where algorithms used for setting prices on goods or services inadvertently use variables that correlate with disability status or result in unequal access to discounts, benefits, or pricing models. For instance, if an algorithmic pricing model disproportionately charges higher prices to users according to their disabled data or inferred that they may have certain relationships with disability, this could constitute discrimination under the ADA. Meanwhile, the ADA requires that all places of public accommodation (including online services) provide equal access to individuals with disabilities. If the implementation of algorithmic pricing creates barriers that prevent individuals with disabilities from equally accessing to goods and services, such pricing practices could potentially violate the ADA. However, one of the key challenges is defining what constitutes discrimination in the context of algorithmic pricing. The ADA was enacted before the digital age, and its provisions do not explicitly address how its anti-discrimination mandates apply to algorithmic processes. In existing algorithmic era, less algorithms will still use sensitive data, such as disabled data, to determine price. One the one hand, it will increase legal risks. On the other hand, they are not necessary to do so, as they only need to obtain users' willingness to pay and the inferred data can largely satisfy this requirement.[46] However, even if sellers are using inferred data, the logic behind is still discriminate certain group of people who may have some kinds of connections with disability.

---

[44] Thorin Klosowski, 'The State of Consumer Data Privacy Laws in the US (And Why It Matters)' (*Wirecutter*, 6 September 2021) <https://www.nytimes.com/wirecutter/blog/state-of-privacy-laws-in-us/> accessed 30 April 2024.
[45] Emmanuel Pernot-Leplay, 'China's Approach on Data Privacy Law: A Third Way Between the US and the E.U.?' (2020) 8 Penn State Journal of Law & International Affairs 49, 55–62.

In the area of financial products, the Gramm-Leach-Bliley Act (GLBA) could also apply in certain scenarios.[47] Known as the Financial Services Modernization Act of 1999, it primarily regulates the collection, disclosure, and protection of consumers' personal financial information by financial institutions. While GLBA is mainly concerned with financial privacy, its provisions can indirectly impact algorithmic pricing, especially within the financial industry. As stipulated in Section 501, financial institutions ensure the security and confidentiality of customer records and information. It requires financial firms to protect against any anticipated threats or hazards to the security or integrity of customer information. Further to this, Section 502 empowers individual with the right to opt out of the sharing of their non-public personal information with non-affiliated third parties. This includes clear notice of the consumer's right to prevent the sharing of personal financial information. In other words, institutions must deliver a privacy notice that outlines their information-sharing practices and the conditions under which they may disclose non-public personal information. In the context of algorithmic pricing, this obligation ensures that consumers are informed about how their data may be used, especially for third parties' sharing. By restricting how and with whom data can be shared, Section 502 indirectly influences the scope of data available for algorithmic pricing. However, it is evident that these stipulations do not address the fairness, bias, or ethical considerations of using algorithms in pricing strategies. They only focus on the security, confidentiality and sharing of personal information without directly tackling how algorithms use this data or the potential for discriminatory outcomes based on the data-driven insights.

In health sector, The Health Insurance Portability and Accountability Act (HIPAA) could also intervene to protect users' privacy and security of certain health information against online algorithmic pricing if it is deployed in relevant context. HIPAA regulates how Protected Health Information (PHI) can be used and disclosed. The PHI is defined in a quite broad way that covers any information, including demographic data, related to an individual's health status, provision of health care, or payment for health care, capable of identifying the individual. The scope of entities covered by HIPAA includes health plans, health care clearinghouses, and health care providers that conduct certain electronic transactions. Unless explicitly permitted or required for treatment, payment, or health care operations, or when authorized by the individual, the use and disclosure of PHI is prohibited. While it seems that HIPAA provides blanket protection for all health information, the protection scope is more narrowly tailored. Firstly, it only covers communication between users and covered entities. In other words, data collected by apps or wearable devices, which monitor health-related information such as physical activity, sleep patterns, or heart rates, is not protected under HIPAA. This means that manufacturers of these devices and apps are free to use the health data they collect in different ways. In the context of algorithmic pricing, for example, a health tracking app could potentially sell user data to insurers firms that use this information to adjust pricing or marketing strategies. Such algorithmic pricing models that incorporate health data not covered by HIPAA can lead to discrimination if not properly regulated. For example, the insurance company can use data from fitness trackers to set premiums, individuals with poor health metrics could end up paying significantly more.

---

[46] Li, 'Affinity-Based Algorithmic Pricing: A Dilemma for EU Data Protection Law' (n 7) 9.
[47] Paul Schwartz and Karl-Nikolaus Peifer, 'Transatlantic Data Privacy Law' (2017) 106 Georgetown Law Journal 148–150.

### 4.2. Regulatory Mechanisms in US Data Privacy Law

As discussed, the remedies and regulatory mechanisms are disparate across different industries against online algorithmic pricing. the FTC Act equips the FTC with authority to pursue legal action against apps and websites that breach their own privacy policies. This power enables the FTC to uphold consumer trust and accountability in digital communications. Additionally, the FTC is tasked with investigating misleading marketing claims related to privacy, including the application of online algorithmic pricing. However, such regimes do not empower individual any rights to protect themselves against algorithmic pricing, which could hinder their privacy and informational autonomy. Such a dependency does not always assure timely or adequate protection of individual privacy interests. The ADA indeed grants individual right to file complaints with the US Department of Justice, which enforces several parts of the ADA. Alternatively, individuals or groups can also file lawsuits against entities that violate their rights under the ADA. However, it is worth noting that several practical barriers to these remedies are making them difficult to apply. For example, lawsuits are costly and the proof burden of discriminatory effect caused by algorithmic pricing is quite heavy because of the inherently opaque nature of many pricing algorithms. In terms of both GLBA and HIPAA, both regimes set forth robust frameworks in their industry. However, none of them provide sufficient direct individual rights against deployers of online algorithmic pricing.

This patchwork of regulatory approaches highlights a critical gap, which is the lack of a unified legal framework that provides individuals with clear, actionable rights against algorithmic pricing across all sectors. As algorithmic pricing becomes more pervasive across different industries, there is a growing need for legislation that not only protects individual data but also empowers people to regain control over how their information is commercially collected, processed and utilised.

## 5. Comparative analysis of data privacy law in the context of online algorithmic pricing: European Union vs United States

As has been analysed above, the legal nature, scope and regulatory measures are different between the EU and US. The following section will compare the regulatory frameworks in both EU and US in the context of online algorithmic pricing.

### 5.1. Legal nature of privacy and personal data in the EU and US

The foundational principles underlying the legal protections for privacy and personal data in the EU and the US are distinctly characterised by their respective historical, cultural, and legal contexts. These differences significantly influence the structure of data privacy laws in both regions. In the EU, the right to privacy and data protection is formally recognised as a fundamental human right. This is explicitly embedded in the Charter of Fundamental Rights of the EU, which includes the right to data protection under Article 8. Therefore, it is evident that such regimes are grounded in broader human rights frameworks, which explicitly require a legal basis for collection and processing personal data; otherwise, such processing is prohibited. Moreover, derived from the concept of privacy, the EU regulatory framework adopt the concept of personal data as regulatable objective. It is a more proactive requirement that places obligations on entities that handle personal data to ensure it is processed fairly, legally, and transparently. By contrast, the US Constitution does not explicitly provide for a right to information privacy, unlike the right to data protection found in the EU.[48] This difference stems from the foundational principles that guided the drafting of the US Constitution, which does not require the government to actively establish conditions that enable the exercise of fundamental rights. Instead, the Constitution establishes a government with strictly limited powers, a design that reflects historical American concerns about the potential for government overreach and oppression. This also represents in the design of whole data privacy framework. The GDPR provides a comprehensive framework that harmonizes data protection laws across all EU member states, ensuring a high level of protection for individuals and stringent obligations on those processing personal data. However, by contrast, the US data privacy law heavily relies on the industrial-based framework and state-led approach.

In the context of online algorithmic pricing, the EU data protection law is notably stronger than that of the US, primarily because the EU treats data protection as a fundamental right. The GDPR not only ensures that personal data is collected and processed under strict conditions but also requires transparency from companies about how they use algorithms and data to set prices. This means businesses must disclose their data processing practices and ensure that their use of algorithms in pricing does not discriminate or unfairly target individuals based on their personal data. By contrast, in the US, data protection is not recognised as a fundamental right at the federal level, resulting in a regulatory environment that varies by state and by industry. This sector-specific and state-led approach leads to inconsistent protections and regulations around the use of algorithmic pricing. This also leads to the different applicable scope of both regimes.

### 5.2. Application scope of the EU and US data privacy laws

Apart from the difference of the legal nature of the EU and US data privacy legislations, the applicable scope of both regions' regulations is quite different, especially in the realm of online algorithmic pricing. For example, the GDPR defines "personal data" broadly to include any information that can identify an individual either directly or indirectly, encompassing a range of data types from basic identity information to digital identifiers like IP addresses.[49] Such extensive coverage ensures that any entity, regardless of industry, that processes personal data is subject to GDPR's stringent directives. For instance, principles of data minimisation and purpose limitation are particularly pertinent to algorithmic pricing, as they require that data collected and processed is strictly necessary for the specified purpose.[50] Moreover, the GDPR mandates robust transparency and gives individuals substantial control over their data with rights to access, rectify, and erase their information, and to object to its processing. This comprehensive approach not only fosters a high level of data protection but also standardizes practices across all member states, offering clarity and consistency in the context of online algorithmic pricing.

---

[48] ibid 11.
[49] Nadezhda Purtova, 'The Law of Everything. Broad Concept of Personal Data and Future of EU Data Protection Law' (2018) 10 Law, Innovation and Technology 40.
[50] Voigt and von dem Bussche (n 22).

Conversely, US privacy regulations are instead dispersed across various statutes. As aforementioned, this segmentation means that personal data protection standards vary significantly from one sector to another, leaving gaps in coverage especially pertinent to newer data practices like algorithmic pricing. Additionally, the use of the term "consumer privacy" in the US reflects a specific conceptual framing of privacy issues that inherently limits the scope of applicable data protection regulations. Primarily used within the context of consumer protection law, it focuses on safeguarding individuals in their capacity as consumers within the marketplace, rather than as citizens with a broad spectrum of privacy rights. In the context of online algorithmic pricing, if narrower interpretation adopted, it is difficult to cover the data collected and processed by deployers of such commercial practice. In contrast to the EU's comprehensive view of data protection as a fundamental right applicable across all aspects of society, rather than treats personal data as a consumer issue. Consequently, while this focus ensures robust protections within consumer transactions, such as preventing deceptive advertising and securing transactional data, it cannot address broader privacy concerns raised by online algorithmic pricing like unfairness, discrimination and informational autonomy.

Such legislative scope of data privacy laws significantly impacts the regulation of online algorithmic pricing. The EU's GDPR offers a comprehensive, uniform framework that applies across all sectors, enhancing transparency and consumer protection but also imposing strict compliance burdens on businesses. In contrast, the US's fragmented, sector-specific approach provides flexibility and is tailored to industry-specific risks but creates gaps by using the consumer privacy definition and complicates compliance efforts, particularly for new technologies and business models that cut across traditional industry boundaries.

### 5.3. Remedies between the US and EU data privacy laws

The regulatory frameworks in the EU and the US differ significantly in their provisions for addressing privacy and personal data concerns, particularly in the context of online algorithmic pricing. The EU GDPR adopts a holistic approach, incorporating both *ex-ante* and *ex-post* mechanisms. There is a stark contrast to the US approach. As aforementioned, the GDPR innovate the regulatory framework by introducing ex-ante measures such as Data Protection Impact Assessments (DPIAs) and Privacy by Design (PbD). These are mandatory for operations that are likely to result in high risks to the rights and freedoms of individuals, including algorithmic pricing. By embedding this *ex-ante* measure into the GDPR framework, it mandates that data protection safeguards be integrated into products and processes from the outset, not as an afterthought. More importantly, the GDPR empowers individuals extensive rights, including the right to access, rectify, erase, and object to the processing of their data, alongside protections against automated decision-making. These rights enable individuals to challenge and seek redress against decisions made by algorithmic pricing models.

In contrast, the US lacks a unified framework and did not grant much individual data rights that victims can exercise to fight against pricing algorithms. Even though the FTC plays a central role in consumer protection, its reach is limited. It cannot enforce privacy practices across all sectors and its actions are often reactive rather than preventative. Moreover, the absence of comprehensive ex-ante measures like DPIAs means that there is no general requirement for companies to assess the systematic privacy risks before implementing algorithmic pricing strategies. Additionally, the lack of expansive individual rights similar to those under the GDPR leaves consumers with fewer tools to challenge unfair or opaque pricing practices. This underscores a critical vulnerability in the US data privacy framework. Without systematic and proactive privacy safeguards, broader protection subjective, and comprehensive individual rights, regulating online algorithmic pricing becomes challenging, potentially leading to inadequate protection of individuals. However, it is also risky to largely broaden the scope of personal data. The GDPR's expansive definition of personal data covers any information that can identify an individual either directly or indirectly. This broad scope is intended to ensure robust protection across all possible data scenarios, but it also introduces a degree of legal uncertainty.[51] The ambiguity can hinder the implementation of the data protection law and innovation. Moreover, the strong individual rights afforded by the GDPR, while empowering consumers, can also be prone to misuse. The right to access, designed to allow individuals to access their personal data, can be exploited to orchestrate data breaches or for other malicious purposes.[52] Even though the GDPR imposes strict obligations and boasts a broad applicability scope, it encounters specific challenges in comprehensively covering affinity data and inference data.[53] However, they are pivotal in the realms of online algorithmic pricing, and present a unique challenge for the GDPR framework.

## 6. The shift paradigm: the new direction to regulate OAP?

With these limitation in mind, the EU not only solely enhancing its framework within the bounds of the GDPR but also pioneering a new regulatory paradigm. Digital Service Act (DSA), Digital Market Act (DMA) and AI Act (AIA) aim to create a safer digital space where the fundamental rights of users are protected and to establish a level playing field for businesses.[54] This shift towards a more horizontal approach is specifically designed to address the evolving concerns of digital technologies and platforms, such as online algorithmic pricing. Firstly, the DSA introduces a new horizontal obligation for digital regulation by implementing measures focused on risk and crisis management and setting specific rules for B2C online marketplaces. It aims to ensure that digital services, including those that use online algorithmic pricing, operate within a framework that promotes transparency, accountability, and fairness. The DSA obligates platforms to manage online risks and crises effectively, ensuring that recommender systems and other algorithms steering online interactions do not harm consumer rights or public safety.

Furthermore, with the concerns of tech giants and large platforms, the DMA targets large online platforms, designated as gatekeepers, and aims to foster a competitive digital market. Similar as the DSA, the DMA emphasizes transparency and the legal basis for personal data collection and processing, highlighting GDPR obligations in the context of digital marketplace. It further reinforces the investigation of profiling by introducing independently audited descriptions. This approach further strengthens the protection of individuals from exploitative pricing algorithms and lays the foundation for fair market practices. Focusing on artificial intelligence, the AI Act introduces specific regulations to certain use of online algorithmic pricing.[55] It introduces a risk-based taxonomy, whereby different use of AI, including online algorithmic pricing, in different sectors has disparate obligations, such as transparency obligations, data governance and human oversight.[56]

---

[51] Li, 'Affinity-Based Algorithmic Pricing: A Dilemma for EU Data Protection Law' (n 7) 9–10.
[52] Wenlong Li and others, 'Mapping the Empirical Literature of the GDPR's (In-)Effectiveness: A Systematic Review' (2025) 57 Computer Law & Security Review 106129.
[53] Li, 'Affinity-Based Algorithmic Pricing: A Dilemma for EU Data Protection Law' (n 7) 11.
[54] European Commission, 'The Digital Services Act Package | Shaping Europe's Digital Future' (2024) <https://digital-strategy.ec.europa.eu/en/policies/digital-services-act-package> accessed 30 April 2024.

From solely relying on the GDPR to embracing broader regulations, this represents a shift in regulatory paradigm within the EU towards a more holistic and horizontal approach to digital regulation, marking a significant evolution in how online algorithmic pricing is governed. Ultimately, this regulatory paradigm shift toward a more foresighted approach is likely to lead to a digital marketplace that is fairer, more competitive, and more innovative. For consumers, this could translate to greater control over their data and more confidence in the fairness of the digital services they use, including how prices are determined algorithmically.

7. Conclusion

To conclude, the comparative analysis above has provided an overview of how EU and US data privacy laws regulate online algorithmic pricing. It demonstrates that the regulatory frameworks in the EU and US are disparate in regulating online algorithmic pricing. The EU not only provides a more comprehensive data protection law, but also turns to a horizontal approach to mitigate the risks of such commercial practices. By contrast, the US regulatory framework is scattered and fragmented, with sector-specific regulations that lead to inconsistencies and gaps in individual protection, particularly in online algorithmic pricing. In the EU, the holistic approach under the GDPR, complemented by the new horizontal regulations like the DSA, DMA, and AI Act, offers robust protections that are uniformly applied across all member states. This regulatory model not only addresses data privacy but also aims to ensure fairness, transparency, and accountability in algorithmic processes, including those used in pricing. The EU's proactive stance in protecting privacy and data protection sets a precedent that could serve as a model for other regions, including the US. Conversely, the US would benefit from a more unified approach to data privacy that transcends the current sector-based framework. While the existing approach offers some degrees of flexibility for firms, it has been proved that this flexibility often results in uneven levels of individual protection and complicates the regulatory environment for businesses operating across different sectors.

However, it does not mean that the EU's approach is a perfect example. The broad scope of what constitutes personal data under EU law introduces its own set of challenges, creating uncertainty that can lead to compliance issues for businesses and enforcement difficulties for regulators and the difficulties for exercising data rights for individuals. This ambiguity is further reflected in the newer horizontal regulatory frameworks such as the DSA and the DMA, where the expansive definition of personal data could complicate the implementation of these laws and make it harder for individuals to exercise their rights effectively. Moreover, the AI Act, which introduces specific regulations for artificial intelligence, including those used in algorithmic pricing, has yet to be accompanied by detailed practical guidelines from authorities. This lack of guidance raises questions about how the AI Act will be enforced in practice and whether businesses will be able to comply effectively. It is worth maintaining a long-term observation of how these new regulations unfold in practice. Assessing how ambiguities are addressed and whether additional clarifications or amendments are made will be crucial in examining the effectiveness of the EU's approach to regulating the digital economy and protecting personal data. Future work should aim to rigorously test these new regulations in various scenarios to evaluate their effectiveness thoroughly. More robust empirical evidence is also desired to compare the difference between "law in books" and "law in action," providing a clearer picture of the actual impact of these regulatory frameworks on digital commerce and consumer protection. This body of empirical research would contribute significantly to the ongoing dialogue about data protection and digital regulation.

---

[55] Zihao Li, 'Why the European AI Act Transparency Obligation Is Insufficient' [2023] Nature Machine Intelligence.
[56] Philipp Hacker, Andreas Engel and Marco Mauer, 'Regulating ChatGPT and Other Large Generative AI Models', *2023 ACM Conference on Fairness, Accountability, and Transparency* (ACM 2023) 1119–1120 <https://dl.acm.org/doi/10.1145/3593013.3594067> accessed 20 November 2023.

**Reference**

Botta M and Wiedemann K, 'To Discriminate or Not to Discriminate? Personalised Pricing in Online Markets as Exploitative Abuse of Dominance' (2020) 50 European Journal of Law and Economics 381

Bu F and others, '"Privacy by Design" Implementation: Information System Engineers' Perspective' (2020) 53 International Journal of Information Management 102124

Capobianco A and Gonzaga P, 'Competition Challenges of Big Data: Algorithmic Collusion, Personalised Pricing and Privacy', *Legal Challenges of Big Data* (Edward Elgar Publishing 2020) <https://www.elgaronline.com/view/edcoll/9781788976213/9781788976213.00008.xml>

Chen J, *Regulating Online Behavioural Advertising Through Data Protection Law* (Edward Elgar Publishing 2021) <https://www.elgaronline.com/view/9781839108297.xml>

Chen L, Mislove A and Wilson C, 'An Empirical Analysis of Algorithmic Pricing on Amazon Marketplace', *Proceedings of the 25th International Conference on World Wide Web* (International World Wide Web Conferences Steering Committee 2016) <https://dl.acm.org/doi/10.1145/2872427.2883089> accessed 8 January 2023

De Bruin R, 'A Comparative Analysis of the EU and U.S. Data Privacy Regimes and the Potential for Convergence' [2022] Hastings Sci. & Tech. L.J. <https://www.ssrn.com/abstract=4251540> accessed 30 April 2024

Esposito F, 'The GDPR Enshrines the Right to the Impersonal Price' (2022) 45 Computer Law & Security Review 105660

European Commission, 'The Digital Services Act Package | Shaping Europe's Digital Future' (2024) <https://digital-strategy.ec.europa.eu/en/policies/digital-services-act-package> accessed 30 April 2024

Hacker P, Engel A and Mauer M, 'Regulating ChatGPT and Other Large Generative AI Models', *2023 ACM Conference on Fairness, Accountability, and Transparency* (ACM 2023) <https://dl.acm.org/doi/10.1145/3593013.3594067> accessed 20 November 2023

Klosowski T, 'The State of Consumer Data Privacy Laws in the US (And Why It Matters)' (*Wirecutter*, 6 September 2021) <https://www.nytimes.com/wirecutter/blog/state-of-privacy-laws-in-us/> accessed 30 April 2024

Li W and others, 'Mapping the Empirical Literature of the GDPR's (In-)Effectiveness: A Systematic Review' (2025) 57 Computer Law & Security Review 106129

Li Z, 'Standardisation of Blockchain and Distributed Ledger Technologies—A Legal Voice from the Data Protection Law Perspective' in Kai Jakobs (ed), *25th EURAS Annual Standardisation Conference – Standards for Digital Transformation: Blockchain and Innovation* (The European Academy for Standardisation 2020)

——, 'Affinity-Based Algorithmic Pricing: A Dilemma for EU Data Protection Law' (2022) 46 Computer Law & Security Review 1

——, 'Why the European AI Act Transparency Obligation Is Insufficient' [2023] Nature Machine Intelligence

——, 'Digital Advertising and EU Digital Regulation', *Digital Advertising Evolution* (Routledge 2024)

Li Z, Yi W and Chen J, 'Accuracy Paradox in Large Language Models: Regulating Hallucination Risks in Generative AI'

Malgieri G, 'Trade Secrets v Personal Data: A Possible Solution for Balancing Rights' (2016) 6 International Data Privacy Law 102

Martin N, 'Uber Charges More If They Think You're Willing To Pay More' (*Forbes*, 2019) <https://www.forbes.com/sites/nicolemartin1/2019/03/30/uber-charges-more-if-they-think-youre-willing-to-pay-more/?sh=7f9256c57365> accessed 3 November 2020

Miller AA, 'What Do We Worry About When We Worry About Price Discrimination? The Law and Ethics of Using Personal Information for Pricing' (2014) 19 Journal of Technology Law and Policy 41

Mittelstadt B, 'From Individual to Group Privacy in Big Data Analytics' (2017) 30 Philosophy and Technology 475

Pernot-Leplay E, 'China's Approach on Data Privacy Law: A Third Way Between the U.S. and the E.U.?' (2020) 8 Penn State Journal of Law & International Affairs 49

Purtova N, 'The Law of Everything. Broad Concept of Personal Data and Future of EU Data Protection Law' (2018) 10 Law, Innovation and Technology 40

Schwartz P and Peifer K-N, 'Transatlantic Data Privacy Law' (2017) 106 Georgetown Law Journal

Seele P and others, 'Mapping the Ethicality of Algorithmic Pricing: A Review of Dynamic and Personalized Pricing' (2021) 170 Journal of Business Ethics 697

Steppe R, 'Online Price Discrimination and Personal Data: A General Data Protection Regulation Perspective' (2017) 33 Computer Law and Security Review 768

Taylor L, Floridi L and Sloot B van der, *Group Privacy: New Challenges of Data Technologies* (Springer International Publishing 2017)

UK Competition and Markets Authority, 'Pricing Algorithms: Economic Working Paper on the Use of Algorithms to Facilitate Collusion and Personalised Pricing' [2018] UK Competition and Markets Authority Working Paper

UK Office of Fair Trading, Coen P and Timan N, 'The Economics of Online Personalised Pricing' [2013] Office of Fair Trading Working Paper 97

Voigt P and von dem Bussche A, *The EU General Data Protection Regulation (GDPR) A Practical Guide* (Springer International Publishing 2017)

Wachter S, 'Affinity Profiling and Discrimination by Association in Online Behavioural Advertising' (2020) 35 Berkeley Technology Law Journal 1

Wachter S and Mittelstadt B, 'A Right to Reasonable Inferences: Re-Thinking Data Protection Law in the Age of Big Data and AI' (2019) 2 Columbia Business Law Review 443

Ward M, 'BBC News | BUSINESS | Amazon's Old Customers "Pay More"' (*BBC News*, 2000) <http://news.bbc.co.uk/1/hi/business/914691.stm> accessed 9 March 2021

Wong B, 'Online Personalised Pricing as Prohibited Automated Decision-Making under Article 22 GDPR: A Sceptical View' (2021) 30 Information & Communications Technology Law 193

Worstall T, 'Why Does Online Pricing Discriminate?' (*Forbes*, 2012) <https://www.forbes.com/sites/timworstall/2012/12/25/why-does-online-pricing-discriminate/> accessed 3 January 2022

Zhang L and others, 'Context-Aware Recommendation System Using Graph-Based Behaviours Analysis' (2021) 30 Journal of Systems Science and Systems Engineering 482

Zhao Z, 'Algorithmic Personalized Pricing with the Right to Explanation' [2023] Journal of Competition Law & Economics <https://academic.oup.com/jcle/advance-article/doi/10.1093/joclec/nhad008/7259590> accessed 17 October 2023

Zuiderveen Borgesius F and Poort J, 'Online Price Discrimination and EU Data Privacy Law' (2017) 40 Journal of Consumer Policy 347

Proposal for a Regulation of the European Parliament and of the Council concerning the respect for private life and the protection of personal data in electronic communications and repealing Directive 2002/58/EC 2017